\documentclass{article}
\usepackage{graphicx}
\usepackage{latexsym}
\newtheorem{theorem}{Theorem}
\newtheorem{lemma}[theorem]{Lemma}
\newtheorem{corollary}[theorem]{Corollary}

\title{Two-Grid Algorithms for Singularly Perturbed Reaction-Diffusion Problems on Layer Adapted
Meshes  
}

\author{Ivanka Tr. Angelova$^*$,
        Lubin G. Vulkov$^+$ \\[0.8ex]
\small {$^*$ University of Ruse, Department of Mathematics,} \\
           \small {Studentska St. 8, 7017 Ruse, Bulgaria} \\
              \small {{\textit{iangelova@ru.acad.bg}}}           
           \\
           \small {$^+$University of Ruse, Department of Applied Mathematics and Statistics,}\\  
           \small {Studentska St. 8, 7017 Ruse, Bulgaria} \\
           \small {{\textit{vulkov@amu.ru.acad.bg}}}}

\date{}
\begin{document}

\maketitle

\begin{abstract}
We propose a new two-grid approach based on Bellman-Kalaba
quasilinearization \cite{cite5} and Axelsson \cite{cite3}-Xu
\cite{cite23} finite element two-grid method for the solution of
singularly perturbed reaction-diffusion equations. The algorithms
involve solving one inexpensive problem on coarse grid and solving
on fine grid one linear problem obtained by quasilinearization of
the differential equation about an interpolant of the computed
solution on the coarse grid. Different meshes (of Bakhvalov,
Shishkin and Vulanovi\'c types) are examined. All the schemes are
uniformly convergent with respect to the small parameter. We show
theoretically and numerically that the global error of the
two-grid method is the same as of the nonlinear problem solved
directly on the fine layer-adapted mesh.
\end{abstract}
%
\section{INTRODUCTION}

We consider the problem

\begin{eqnarray}\label{eq1}
&- \varepsilon^{2} u'' + f(x,u) = 0 , \;x \in \Omega \equiv (0,1), \\
& u(0) = u (1) = 0, \nonumber
\end{eqnarray}
where $ \varepsilon $ is a small perturbation parameter, $ 0 <
\varepsilon \ll 1 $. We assume (see Lemma \ref{lem1} and Theorems
\ref{th2}, \ref{th3}) that the function $f$ has the continuous
derivatives:
\begin{equation}\label{eq2}
\frac{\partial ^{i+j} f(x,u)}{\partial x^i \partial u^j},\;0\leq
i+j \leq 4,\;\;0\leq i \leq 3,\;\;0\leq j \leq 4,\;(x,u)\in  (
\overline{\Omega} \times R ),
\end{equation}
and
\begin{equation}\label{eq3}
f_{u} (x,u) \geq c_{0}^{2} > 0 , \; (x,u) \in ( \overline{\Omega}
\times R ).
\end{equation}
The condition (\ref{eq3}) is the standard stability condition,
which implies that both (\ref{eq1}) and the reduced problem
$f(x,u) = 0$ have unique smooth solutions $ u_{\varepsilon }\in
C^4 (\overline{\Omega})$ and $ u_{0} $, respectively.

It is shown theoretically and experimentally in \cite{cite6} that
there exists no finite difference scheme (or finite element
approximation) of (\ref{eq1}) on standard meshes whose solution
can be guaranteed to converge to the solution $u$ in the maximum
norm, uniformly with respect to the perturbation parameter
$\varepsilon$.

Nowadays, two basic types of non-equidistant (layer adapted)
meshes, suggested by Bakhvalov and Shishkin, are used for solving
singularly perturbed problems \cite{cite11,cite13}. An explicit
mesh construction method to solve a singularly perturbed problem
of type (\ref{eq1}) was used first by Bakhvalov \cite{cite4},
where he obtained the special discretization mesh $ w_{h} = \{
x_{i} = \lambda ( i / n ) : i=0,1, \dots n \}$, $ h = 1/n $, where
by $\lambda$ is the mesh generating function that consists of
three parts: $ \lambda_{1},\;\lambda_{2}$ and $\lambda_{3}$. The
functions $ \lambda_{1} $ and $ \lambda_{3} $ generate the mesh
points in the boundary layers in the neighborhood of $x=0$ and
$x=1$ respectively. The function $\lambda_{2}$ generates the mesh
points outside the boundary layers and it is a tangent line to
both $\lambda_{1} $ and  $\lambda_{3} $, and $
\lambda_{2}(0.5)=0.5$.

A much simpler mesh was constructed by Shishkin, see
\cite{cite11,cite13}, but many difference schemes applied on
Bakhalov's mesh show better results.

Not only to simplify Bakhvalov's mesh but also to increase the
density of mesh points in the boundary layers, Vulanovi\'c
modifies the previously known mesh generating functions
\cite{cite16}, see also \cite{cite7,cite10,cite13,cite18,cite19}.

There is a wide range of publications that deal with layer adapted
meshes. In the monograph \cite{cite10} an extensive review is
given.

Currently there is  considerable interest in the construction of
high-order approximations to singularly perturbed problems,
\cite{cite10,cite13,cite14,cite15,cite17,cite18,cite19}. But such
constructions often lead to an extension of the stencil or to
discretizations that are not inverse monotone \cite{cite10}.
Another way to increase the accuracy of the numerical solution to
singularly perturbed problems is the use of Richardson
extrapolation \cite{cite10,cite13,cite14}. However the Richardson
procedure requires solution of systems of nonlinear algebraic
equations on each of the nested meshes \cite{cite14}.

The main objective of this paper is to present  two-grid
algorithms using standard difference approximation on different
adaptive meshes for the boundary value problem (\ref{eq1}). The
two-grid method used for high-order and time-effective
computations was first introduced by Axellson \cite{cite3} and Xu
\cite{cite23} independently from each other. It was further
investigated by many other authors and for many problems for
instance, for elliptic, parabolic and Stokes-Darcy equations, see
\cite{cite12} and the references therein. Note that the error
estimates in these papers are in
 {\it{weak (Sobolev-type) }} discrete norms. In comparison, our error estimates below are
in the {\it{maximum }} norm. This norm is sufficiently strong to
capture layers and hence seems most appropriate for singularly
perturbed problems.

The rest of the paper is organized as follows. In Section 2 we
introduce three meshes. In Section 3 we describe a Newton-Bellman
$\&$ Kalaba linearization process \cite{cite5,cite9} for the
continuous problem (\ref{eq1})-(\ref{eq3}) in order not only to
prove uniform convergence of the difference scheme but mainly to
obtain the estimate (\ref{eq15}) which plays a key role in the
analysis of the two-grid algorithms in the next section. In
Section 4 we describe the two-grid algorithms (TGAs) and provide
error estimates for the difference scheme discretization of the
TGAs, Theorem \ref{th3}. Section 5 includes numerical results that
illustrate the theoretical estimates. Finally, conclusions and
directions for future work are presented.

Although our theoretical results will be presented in a model,
one-dimensio\-nal classical situation, the algorithm possesses a
wider generality, see test Example 2 in Section 5. Also, there are
many interesting and relevant boundary value problems of type
(\ref{eq1}), for which the condition (\ref{eq3}) is not satisfied,
see \cite{cite25,cite8}, and for which our computational
techniques still work well.

{\bf {Notation }} We define a norm of a continuous function $f(x)$
  as $ \Vert f \Vert =\max\limits_{x \in \bar{\Omega}}|f(x)|.$
Let $w_{h} = \{ 0  < x_{1} < \dots < x_{n-1} < 1 \},\;
x_{0}=0,\;x_{n} =1,\;\overline{w}_{h}=w_{h}\cup \{x_0\}\cup
\{x_n\}$ and $ h_{i} = x_{i} - x_{i-1},\; \hbar_{i} = 0.5 ( h_{i}
+ h_{i+1} )$. For a mesh function $y$, we introduce the standard
finite difference approximations to the first and second
derivatives, \cite{cite11,cite13}:
$$ y_{  { \overline {x}},i   } =  (y_{i} - y_{i-1} ) / h_{i} ,
\;  y_{x,i} =  y_{  { \overline {x}},i+1   } , \; y_{  {\widehat
{x}} , i   }=(y_{i} - y_{i-1} ) / \hbar_{i} ,$$
$$  y_{  { \overline {x}} { \widehat {x}},i} =
\frac{1}{ \hbar_{i}   }\left(\frac{  y_{i+1} - y_{i} } { h_{i+1} }
-\frac{  y_{i} - y_{i-1} } { h_{i} } \right).$$ The discrete
maximum norm is defined by $\Vert y \Vert= \Vert y \Vert_{h}  =
\max\limits_{ 0 \leq  i \leq n } \vert y_{i} \vert.$  Throughout
this paper $C$ and $C_i,\ i\ge 0$, denote  positive constants
independent on $N$ (the number of coarse mesh nodes, respective
mesh step $H$), $n$ (the number of fine mesh nodes, respective
mesh step $h$) and $ \varepsilon $.

\section{THE LINEAR PROBLEM ON ADAPTIVE MESHES}

For a given integer $n$,  on $\overline{\Omega} = [0,1]$ we
introduce the special mesh $\overline{w}_{h}$ with $ x_{i} =
\lambda ( i /n) , \;  i=0,1, \dots , n .$ Following
\cite{cite7,cite10,cite13} we present all such meshes by their
mesh-generating functions. Bakhvalov's mesh-generating function is
given by
$$
 \lambda (t) =
   \left\{
      \begin{array}{ll}
       \phi (t) : = a \varepsilon \ln \frac{q}{q-t} , & t \in [0, \alpha ] , \\
       \phi ( \alpha ) + \phi ' ( \alpha ) (t- \alpha ) , & t \in [ \alpha , 0.5 ] , \\
       1 - \lambda (1-t)  , & t \in [0.5 , 1 ],
       \end{array}
   \right.
$$
where $a$ and $q$ are constants, independent of $\varepsilon $,
such that
\begin{equation}\label{eq4}
q \in ( 0 , 0.5),\; a \in \left(0,\frac{q}{\varepsilon }\right).
\end{equation}
Here $\alpha $ is the abscissa of the contact point of the tangent
line from $( 0.5 , 0.5 )$ to $ \phi(t)$. The generated mesh will
be called $B$ - mesh.

Shishkin's mesh (S-mesh) is a piecewise equidistant and
consequently much simpler than the mesh above. The generating
function for this mesh is
$$
 \lambda (t) =
   \left\{ \begin{array}{ll}
       4 \alpha t , & t \in [0, \alpha ] , \\
        \alpha  + 2  (1 - 2 \alpha ) ( t-0.25)  , & t \in [ \alpha , 0.5 ] , \\
       1 - \lambda (1-t)  , & t \in [0.5 , 1 ]
      \end{array}\right.
$$
with
$$
\alpha = \min \{ 1/4 , 2 \gamma_{0}^{-1}
   \varepsilon \ln N \} , \;
   \gamma_{0} = \min \{ c_{0} ,1 \}.
$$

Vulanovi\'c, see \cite{cite7,cite16,cite18},  has shown that
$\lambda $ does not need to be a logarithmic function. A class of
suitable mesh generating functions was given and it includes
functions of a much simpler rational form. From those functions we
select the following one
$$
 \lambda (t) =
   \left\{\begin{array}{ll}
       \mu(t) : =  \frac{ a \varepsilon t}{q-t} , & t \in [0, \alpha ] , \\
       \mu ( \alpha ) + \mu ' ( \alpha ) (t- \alpha ) , & t \in [ \alpha , 0.5 ] , \\
       1 - \lambda (1-t)  , & t \in [0.5 , 1 ],
       \end{array}\right.
$$
where $q$ and $a$ satisfy the conditions (\ref{eq4}) and the
parameter $ \alpha $ has the same meaning as in the Bakhvalov
mesh, but it can be explicitly calculated,

$$ \alpha =\frac{ q -{\sqrt { \varepsilon a q (1-2q+ 2 \varepsilon a)
}}} {  1 + 2 \varepsilon a }.$$ This mesh will be called V-mesh.

In the next section we shall develop a linearization procedure for
the solution of problem (\ref{eq1})-(\ref{eq3}). At each iteration
we solve linear two-point boundary value problems of the following
type:
\begin{eqnarray}\label{eq5}
& - \varepsilon^{2} u'' + b( x,\varepsilon ) u = g ( x,
\varepsilon
) , \; x \in \Omega  , \\
& u(0)=0 , \; u(1)=0. \nonumber
\end{eqnarray}

The functions $b(x,\varepsilon )$ and $g(x ,\varepsilon )$ are
assumed to be in $C^3(\overline{\Omega} )$ for a fixed parameter
$\varepsilon \in [0,1]$. On each of the meshes above we shall
consider the classical three-point difference scheme for the
problem (\ref{eq5}):

\begin{eqnarray}\label{eq6}
& - \varepsilon ^2 y_{{\overline {x}} {\widehat {x}},i}+
b(x_{i},\varepsilon ) y_{i} = g(x_{i},\varepsilon), \\
& y_{0} = 0 , \; y_{n} = 0. \nonumber
\end{eqnarray}

Using the discrete Green's function method, optimal convergence
results for the discrete solutions of  two-point boundary value
problems can be obtained, see \cite{cite24,cite25,cite10,cite13}
and references therein. On the base of classical results for the
case $b( x,\varepsilon )=b(x),\;g( x,\varepsilon )=g(x)$, see
\cite{cite4,cite11,cite13,cite25} one can prove easily the
following theorem:

\begin{theorem}\label{th1} Let $b,\; g$ have continuous derivatives with respect to $x$ up to order three that are uniformly
bounded with respect to $\varepsilon \in [0,1]$ and
$b(x,\varepsilon)\geq \beta
> 0$ for all $(x,\varepsilon ) \in (\overline{\Omega} \times
[0,1])$. If $u$ is the solution of problem (\ref{eq5}) and $y$ of
problem (\ref{eq6}), then for the error of the difference scheme
(\ref{eq6}) the following estimate holds:
\begin{equation}\label{eq7}
 \Vert u - y  \Vert \leq C n^{-2} \ln^{k} n, \; \; \left\{
   \begin{array}{l}
     k=2 \; on \; S-mesh , \\
     k=0  \; on \;  B \; and \; V-meshes .  \\
   \end{array} \right.
\end{equation}
\end{theorem}

\section{UNIFORM CONVERGENCE VIA NEWTON'S LINEARIZATION }

On the mesh $\overline{w}_{h}$ the scheme (\ref{eq6}) for
(\ref{eq1})-(\ref{eq3}) is defined as follows:
\begin{equation}\label{eq8}
- \varepsilon^{2} y_{{ \overline {  x }} { \widehat { x }} , i } +
f (x_{i}, y_{i}) = 0 , \; i=1, \dots , n-1,
\end{equation}
$$y_{0} = 0 , \; y_{n} = 0. $$

The uniform convergence for this scheme was analyzed in many
papers \cite{cite13,cite16,cite17}. We shall also address this
 using a linearization Newton-Bellman$\&$Kallaba procedure
\cite{cite5,cite9}. In a natural way this will lead us to the idea
of the two-grid method.

The following assertion is well known and often used to prove
uniform convergence of the scheme (\ref{eq8}) (see for example
\cite{cite13,cite17,cite18}).

\begin{lemma}\label{lem1}
Let the conditions (\ref{eq2}), (\ref{eq3}) be satisfied. Then:

{ \bf { a) }} the problem (\ref{eq1})-(\ref{eq3}) has unique
solution $u\in C^{4 }(\overline{\Omega} )$.

{\bf {b)}} the solution $u$ satisfies the estimates:
\begin{eqnarray*}
\Vert u^{(j)} \Vert &\leq & C(1+\varepsilon^{-j}(\exp(-c_{0}x /
\varepsilon )+ \exp (- c_{0} (1-x)/ \varepsilon ))
\end{eqnarray*}
for $ j=1,2,3. $ The estimate for $j=0 $ looks as follows:
$$
\Vert u \Vert  \leq   c_{0}^{-2}\Vert f(x,0) \Vert .
$$
\end{lemma}

We use the quasilinearization of Bellman$\&$Kalaba for studying
the problem (\ref{eq1})-(\ref{eq3}):
\begin{eqnarray}\label{eq9}
&& L^{m} u^{ (m+1)  } \equiv - \varepsilon ^2\frac{ d^{2} u^{
(m+1)}}{d x^{2}}+
f_{u}' (x,u^{(m)})u^{(m+1)}=-f_{u}(x,u^{(m)})+f_{u}'(x,u^{(m)})u^{(m)} \nonumber \\
&& u^{(m+1)}(0)=0,\;u^{(m+1)}(1)=0,\;m=0,1,2 \dots  \bullet
\end{eqnarray}

Let us  first establish convergence of the linearization process.
Suppose that
\begin{equation}\label{eq10}
\Vert u - u^{ (0)  } \Vert \leq \rho .
\end{equation}
Let
$$\theta = \max_{x \in \overline{\Omega} ,\vert \xi \vert \leq l + 2\rho }
\Vert f_{ uu }''(x,\xi )\Vert .$$

\begin{lemma}\label{lem2}
Assume that $c_0 ^{-2}\theta \rho <1$. Then
\begin{equation}\label{eq11}
\|u^{(m)}-u\|\leq c_0 ^2\theta ^{-1} ( c_0 ^{-2}\theta
\rho )^{2^{m}},\; m=0,1,2,\cdots \bullet
\end{equation}

Also, if the function $u^0(x)$ is in $C^3$ and satisfies estimates
of type b) for $j=0,1$ in Lemma \ref{lem1}, then for the solution
$u^{m+1}$ assertions of type a), b) in Lemma \ref{lem1} hold.
\end{lemma}

\noindent {\textit{Proof.}}  The boundary value problem for $v=u^{(m+1)}-u$ reads as follows:

\begin{equation}\label{eq12}
L^{m} v=F^{(m)} (x) ,\;v(0)=0,\;v(1)=0,
\end{equation}
where
$$F^{(m)}=f(x,u^{(m)})-f(x,u)+f'_{u}(x,u^{(m)})u^{(m)}-f'_{u}(x,u^{(m)})u.$$
We will prove by induction that for all $s \geq
0,\;\|u^{(s)}-u\|\leq \rho.$ For $k=0$ this inequality is obvious.
Suppose that it holds for $s=m$. Using the mean value theorem, we
easily obtain $\|F^{(m)})\| \leq \theta \|u^{(m)} -u\|^2$. The
maximum principle applied to problem (\ref{eq12}) implies:
\begin{equation}\label{eq13}
 \|u^{(m+1)} -u\|\leq c_{0}^{-2}\theta \|u^{(m)} -u\|^2.
\end{equation}
By the assumptions $\|u^{(m)} -u\| \leq \rho $ and
$c_{0}^{-2}\theta \rho <1$ we reach the next step of the
induction. So that for all $m \geq 0$ we have $\|u^{(m)}-u\| \leq
\rho $. Therefore, (\ref{eq13}) holds for all $m\geq 0$ and this
implies (\ref{eq11}).\hfill  $\Box$

Furthermore, after appropriate choice of $u^0$ ($u^0=y^{I(x)}_H$)
in Algorithm 1, and taking into account of assumptions
(\ref{eq2}), (\ref{eq3}) the equation (\ref{eq9})  at $m=0$ takes
the standard form studied in \cite{cite11,cite13}. Then, by
induction one can easily prove the second part in the formulation
of the present Lemma.

Let us consider the finite-difference analogue of the iterative
process (\ref{eq9}):
\begin{eqnarray}\label{eq14}
&&L_{h} y_{i}^{(m+1)}=\varepsilon ^2 y_{\bar{x} \hat{x},i}^{(m+1)} +
f_{u}'(x_{i},y_{i}^{(m)})y_{i}^{(m+1)} \nonumber \\
&&=-f(x_{i},y_{i}^{(m)})+f_{u}'(x_{i},y_{i}^{(m)})y_{i}^{(m)},
\;i=1,\cdots , n-1,\\
&&y_{0}^{(m+1)}=0,\;y_{N}^{(m+1)}=0,\;y_{i}^{(0)}=u^{(0)}(x_{i}),\;i=1,\cdots
, n-1,\;m=0,1,2,\cdots  \bullet \nonumber
\end{eqnarray}

\begin{theorem}\label{th2} Under the conditions (\ref{eq2}), (\ref{eq3}) there exist constants $n_0$ and $\rho_0$,
independent of $\varepsilon$, such that if $n \geq n_0$ and
$\|u^{(0)}-u\| \leq \rho \leq \rho _0$, then the following
estimate holds:
\begin{eqnarray}\label{eq15}
&& \|y^{(m)}-u\|\leq C \left[n^{ -2 }  \ln^{k} n +(c_0 ^{-2}
\theta \rho )^{2^{m}}\right],\; \\
&&   \left\{
    \begin{array}{l}
     k=2 \; on \; S-mesh, \\
     k=0 \; on \; B\; and\; V-meshes.
     \end{array}
     \right. \;
m=0,1,2,\cdots \bullet \nonumber
\end{eqnarray}
\end{theorem}

\noindent {\textit{Proof.}} Assuming that $c_0 ^{-2} \theta \rho <1$, we
introduce the auxiliary iterative process:
\begin{eqnarray*}
&&-\varepsilon ^2 \tilde{y}_{\bar{x} \hat{x},i}^{(m+1)} +
f_{u}'(x_{i},u^{(m)}(x_{i}))\tilde{y}_{i}^{(m+1)}  \\
&&=-f(x_{i},u^{(m)}(x_{i}))+f_{u}'(x_{i},u^{(m)}(x_{i}))u^{(m)}(x_{i}),
\;i=1,\cdots n-1,\\
&&\tilde{y}_{0}^{(m+1)}=0,\;\tilde{y}_{n}^{(m+1)}=0,
\;m=0,1,2,\cdots  \bullet
\end{eqnarray*}

An application of  Theorem \ref{th1} provides the estimate
\begin{eqnarray}\label{eq16}
\|\tilde{y}^{(m+1)}-u^{(m+1)}\|\leq C  n^{ -2 } \ln^{k} n , \\
\left\{
    \begin{array}{l}
     k=2 \; on \; S-mesh, \\
     k=0 \; on \; B\; and \;V -meshes,
     \end{array}
     \right. \;
\;m=0,1,2,\cdots  \bullet \nonumber
\end{eqnarray}
Now we will show that $\|y^{(m+1)}-\tilde{y}^{(m+1)}\|$.
$v^{(m+1)}=y^{(m+1)}-\tilde{y}^{(m+1)}$ satisfies the difference
problem:
\begin{equation}\label{eq17}
L_{h}v_{i}^{(m+1)}=F_{i}^{(m)},\;i=1,\cdots
,n-1,\;v_{0}^{(m+1)}=0,\;v_{n}^{(m+1)} = 0,
\end{equation}
where for $F_{i}^{(m)}$ we have the representation
\begin{eqnarray*}
&&F_{i}^{(m)}=-(y_{i}^{(m)}-u^{(m)}(x_{i}))(f_{uu}(x_{i},\xi_{i}^{(4)})(\xi
_{i}^{(1)}-\xi _{i}^{(3)}) \\
&&+\tilde{y}_{i} ^{(m+1)} (f_{uu} (x_{i} ,\xi_{i}^{(2)}) )
-f_{uu}(x_{i},\xi_{i}^{(3)})+f_{uu}(x_{i},\xi_{i}^{(3)})(\tilde{y}_{i}
^{(m+1)}-\xi_{i}^{(3)})),
\end{eqnarray*}
and all $\xi_{i}^{(j)}$  $ j=1,2,3,4 $ lie between $y_{i}^{(m)}$
and $u ^{(m)}(x_{i})$. We estimate $\tilde{y}_{i}
^{(m+1)}-\xi_{i}^{(3)}$ using
$$
{ \widetilde {y}}^{(m+1)}_{i}  - \xi_{i}^{(3)} =
( { \widetilde {y}}^{(m+1)}_{i}- u^{(m+1)} (x_{i})  ) ) +
( u^{(m+1)}(x_{i})  - u^{(m)}(x_{i})  )
+ ( u^{(m)} (x_{i})  - \xi^{(3)}_{i } ) .
$$

Starting from (\ref{eq17}) and using (\ref{eq11}), (\ref{eq16}),
by an analogical way as in Lemma \ref{lem2}, one can show that for
sufficiently large $n_{0} $ and small $ \rho_{0} $ the following
estimate holds $ \Vert y^{(m)} - u^{(m)} \Vert \leq \rho $ for all
$m \geq 0 $. Therefore, $\xi_{i}^{(j)} , \;j=1,2,3,4 $ are bounded
and there exist constants $ C_{3} , C_{4}, C_{5} $ such that:
\begin{equation} \label{eq18}
     \Vert F^{(m)} \Vert \leq ( C_{3} \Vert y^{(m)} - u^{(m)} \Vert + C_{4}
     \Vert  u^{(m+1)} - u^{(m)} \Vert +C_{5} n^{-2} \ln^{k} n ) \Vert
     y^{(m)} - u^{(m)} \Vert .
\end{equation}

We used that the continuous functions $f_{uu},\;f_{uuu}$ with
given arguments are bounded.

Applying the maximum principle to problem (\ref{eq17}) and using
(\ref{eq16}), (\ref{eq18}), we obtain
\begin{equation} \label{eq19}
\Vert y^{(m+1)} - u^{(m+1)}  \Vert \leq
\end{equation}
$$ c_{0}^{-1} ( C_{3} \Vert y^{(m)} - u^{(m)} \Vert + C_{4}
\Vert  u^{(m+1)} - u^{(m)} \Vert +C_{5} h^{2} )  \Vert y^{(m)} -
u^{(m)} \Vert + C_{2} n^{-2} \ln^{k} n. $$

We make another restriction on $n_{0} $ and $\rho_{0}$:
$$n^{-2} \ln^{k} n \leq c_{0}^{-2} / ( 6 C_{5} ) , \; \rho \leq \min ( \alpha / ( 6 C_{4} ),
\alpha / ( 6 C_{3} )). $$

Now, in view of $ \Vert y^{(m)} - u  \Vert \leq \rho $, it follows
from (\ref{eq19}), that
$$ \Vert y^{(m+1)} - u^{(m+1)}  \Vert \leq 0.5 \Vert y^{(m)} - u \Vert + C_{2} n^{-2} \ln^{k} n , \; m \geq 0.$$

Hence,
$$\Vert y^{(m)} - u^{(m)}  \Vert \leq  C_{2} n^{-2} \ln^{k} n , \; m \geq 0 .$$

Finally, (\ref{eq11}) implies (\ref{eq15}). \hfill $\Box$

Now, we are in a position to prove that the scheme (\ref{eq8}) is
uniformly convergent with respect to $ \varepsilon $.

\begin{corollary}\label{corollary1}
Let $u$ be the solution of problem (\ref{eq1})-(\ref{eq3}) and $y$
f the discrete problem (\ref{eq8}). Then he following estimate of
type (\ref{eq7}) holds true
\begin{equation}\label{eq20}
 \Vert u - y  \Vert \leq C n^{-2} \ln^{k} n, \; \; \left\{
   \begin{array}{l}
     k=2 \; on \; S-mesh , \\
     k=0  \; on \;  B \; and \; V - meshes. \\
   \end{array} \right.
\end{equation}
\end{corollary}

\noindent \textit{Proof.}
Let us chose $ n_{0} ,\; \rho_{0} $ in agreement with the
requirement of Theorem \ref{th2}: $ n \leq n_{0} , \; \rho \leq
\rho_{0} $. Similarly as in Theorem \ref{th2} one can prove that
$$
\|y^{(m)}-u\|\leq c_{0}^{2} \theta_{0}^{-1} (c_0 ^{-2} \theta \Vert y^{(0)} - y \Vert
)^{2^{m}} , \; m=0,1,2, \dots  \bullet
$$

Therefore
$$
y^{(m)} \to y , \; \mbox{as} \; m \to \infty ,
$$
if $ \rho = c_0 ^{-2} \theta \Vert y^{(0)} - y \Vert < 1 $ . Let $
m \to \infty $ , then from (\ref{eq15}) we get the required
estimate.

To complete the proof let us consider the case $ n < n_{0} $. The
maximum principle implies $ \Vert y \Vert \leq l $  ( $l$
corresponds to those in Lemma 1,b). Hence
$$\Vert y-u \Vert \leq \Vert y \Vert + \Vert u \Vert \leq 2 l = C n^{-2} \ln^{k} n , \;
C = 2l
  \left\{
   \begin{array}{l}
     \frac{  n^{2}   }{  \ln ^2 n  } \; on \; S-mesh , \\ \\
     n^{2} \; on \;  B\;and \;V - meshes. \\
   \end{array} \right. \hfill \Box
$$

\section{TWO-GRID  ALGORITHMS}
In this section we propose two-grid algorithms \textit{based on
the estimate (\ref{eq15})}. For this we introduce the fine grid $
\overline{w}_{h}$ with $ x_{i} = \lambda ( i / n ) , \; i=1,
\dots , n,\;\mbox{ and }n=N^r$, where $ r> 1 $ is a real number
that will be chosen later, see Theorem \ref{th3} and comments on
Table 6 in Section 5.

A nice property of Shishkin (Bakhalov and Vulanovi\'c)
\cite{cite7,cite13,cite22} meshes is parameter uniform
interpolation. On the coarse grid
$$w_{H} = \{ 0 = X_{0} < X_{1} < \dots < X_{N-1} < X_{N} =1 \},
$$
define the linear interpolation of the solution of the discrete
problem (\ref{eq8})
$$
y_{N}^{I} (x) = \sum_{  i=0  }^{N} y_{i} \phi_{i} (x) ,
$$
where $ \phi_{i} (x) $ is the standard piecewise linear basis
function associated with the interval $ [ X_{i-1} , X_{i+1} ] $ .
For the interpolant $ y_{N}^{I} (x) $ the following estimate holds
$$\Vert y_{N}^{I} (x) - u \Vert \leq
\Vert u^{I} - u \Vert + \Vert y_{N}^{I} - u^{I} \Vert $$
$$\leq  C N^{-2} \ln^{k} N, \;
  \left\{
   \begin{array}{l}
     k=2 \; on \; S-mesh , \\
     k=0   \; on \; B \; and \;  V -  meshes, \\
   \end{array} \right.$$
where $ u^{I} = u^{I} (x) $ is the interpolant of the continuous
solution $u$. If in the iterative process (\ref{eq9}) one takes
$m=1$ and the initial guess $ u^{(0)} (x) = y_{N}^{I} (x) $ , then
in (15) we will have
$$\left(c_0 ^{-2}\theta \rho \right)^2= C N^{-4} \ln^{2k} N, \;
\left\{
   \begin{array}{l}
     k=2 \; on \; S-mesh , \\
     k=0   \; on \; B \; and \;  V - meshes.
\end{array} \right.$$
Then
$$\Vert y^{1}- u \Vert _{h}\leq C\left[ N^{-2r} \ln^{k} N + N^{-4} \ln^{2k} N\right], \; \;
  \left\{
   \begin{array}{l}
     k=2 \; on \; S-mesh , \\
     k=0   \; on \; B \; and \;  V - meshes, \\
   \end{array} \right.$$
Our first algorithm reads as follows.

{\bf {Algorithm 1}}

{\bf {Step 1.}} Solve the discrete problem (\ref{eq8}) on the
coarse grid $ w_{H}$ and then perform a linear interpolation to
obtain the function $ y_{H}^{I} (x)$ defined in the domain
$\overline {\Omega  } = [0,1]$.

{\bf {Step 2.}} Solve the linear discrete problem
\begin{eqnarray*}
& - \varepsilon^{2} y_{\overline {x} { \hat {x} },i} + f_{u}'
(x_{i} ,y_{N}^{I}(x)) y_{i} \\
&  = f_{u}' ( x_{i} , y_{N}^{I}(x) ) y_{i}-f ( x_{i}
,y_{H}^{I}(x_{i})), \;i=1 , \dots , n-1,\\
& y_{0} = 0 , \; y_{n} = 0
\end{eqnarray*}
to find the fine mesh numerical solution  $ y^{h} $.

{\bf {Step 3.}} Interpolate $ y^{h} $ to obtain $ u_{h,H}^{I} (x)
, \; x \in \overline { \Omega } $.

The next theorem gives the main theoretical result of the present
paper.
\begin{theorem}\label{th3}
Let the conditions (\ref{eq2}), (\ref{eq3}) hold and $n=N^2$, i.e.
$r=2$. Then the following error estimate holds true:
\begin{equation}\label{eq21}
 \Vert u_{h,H}^{I} - u  \Vert _{H}= C N^{-4} \ln^{2k} N, \; \; \left\{
   \begin{array}{l}
     k=2 \; on \; S-mesh , \\
     k=0  \; on \;  B \; and \; V-meshes . \\
   \end{array} \right.
\end{equation}
\end{theorem}

It is clear that we can repeat Algorithm 1 to obtain, on  the fine
mesh $ w_{h} $, with $ n=N^{4} $, the accuracy
$$ C N^{-8} \ln^{4k} N, \; \; \left\{
   \begin{array}{l}
     k=2 \; on \; S-mesh , \\
     k=0  \; on \;  B \; and \; V-meshes . \\
   \end{array} \right.
$$

{\bf {Algorithm 2}}

{\bf {Step 1.}} For $ m=0 $ do step 1 of Algorithm 1.

{\bf {Step 2.}}  For $ m=1,2, \dots $ repeat step 2 of Algorithm 1
with final mesh step corresponding to $ n=N^{2^m} $ and $
y_{N}^{I} (x) := u^{I}_{ h^{m-1} , H} (x)$.

The rate of convergence is the same as in Algorithm 1. However,
there is a significant decrease in the computational cost.

{\bf {Remark 1.}} The formulas  of two-grid Axelsson-Xu type
algorithms \cite{cite3,cite12,cite23} involve the second
derivative $f_{uu}$, while our Algorithms 1, 2 are free of  the
second derivative.

\section{NUMERICAL RESULTS}
In this section we discuss numerical results for a set of
computational experiments associated with the TGAs.

\begin{centering}
\begin{table} \label{table1}
\caption{Points in boundary layers (\%) of coarse grid (N) and
fine grid (n).}
\begin{tabular}{ccccccc}
\hline\noalign{\smallskip}
\hline  & N & 8 & 16& 32& 64 \\
\cline{2-6}  & n & 64 & 256 & 1024  &  4096  \\
\noalign{\smallskip}\hline\noalign{\smallskip}
\hline S - mesh  & Step 1 & 25  & 12.5 &  12.5 &  6.25 \\
\cline{2-6}& Step 2& 6.25 & 4.69 & 3.71 & 3.03 \\
\hline V - mesh &Step 1 & 50 & 50 & 43.75 & 40.63 \\
\cline{2-6} &Step 2 & 40.63 &  40.63 &  40.04 & 40.04 \\
\hline B - mesh &Step 1& 25 & 25 & 18.75 & 18.75 \\
\cline{2-6} &Step 2& 18.75 & 17.97 & 17.77 & 17.72 \\
\noalign{\smallskip}\hline
\end{tabular}
\end{table}
\end{centering}

In order to emphasize the difference between meshes we present
Table 1, where the percentage of the number of mesh points in the
boundary layers, i.e. in $[0, \varepsilon ]\cup [ 1-\varepsilon, 1
]$, is given, with $q=0.4,\;a=1$ and $\varepsilon = 2^{-8}$.

\noindent \textbf{Example 1.} \label{exa1}
We first consider the test problem \cite{cite17}
$$
-\varepsilon ^2 u'' +\frac{u-1}{2-u}+f(x)=0, \; u(0)=u(0)=0,
$$
where $f(x)$ is chosen so that the exact solution is
$$
u_{\varepsilon}(x)=1-\frac{\exp(-x/\varepsilon)+\exp(-(1-x)/\varepsilon)}{1+\exp(-1/\varepsilon)}.
$$

The tables below present the errors
$$E_{N}=\|u_{\varepsilon}-y\| ,$$
where $y$ is the numerical solution on a mesh with N mesh steps.
Also, we calculated numerical orders of convergence by formula
$$O_{N}=\frac{\ln E_{N}-\ln E_{2N}}{\ln 2}.$$

Also, we introduce $O_n = O_N(TGAs(N^2))$, i.e. the order of
convergence of TGAs with $r=2$, and $O_{N^r} = O_N(TGA(N^r))$, the
order of convergence of TGAs with $n=N^r$.

Numerical results are presented in Tables 2, 3, 4 which validate
the theoretical ones established in Theorem 2 and Corollary
\ref{corollary1}. It is interesting to discuss the computations
for small $\varepsilon$. In this case the methods are uniformly
convergent, the errors stabilize for each $N$ as $\varepsilon
\rightarrow 0$. See the results in the Tables for $\varepsilon =
10 ^{-2},\;\varepsilon = 10 ^{-4}$. So, we will discuss the
correspondent rows in the tables for $\varepsilon = 10 ^{-2}$. For
example the maximum error at the Step1 for $N=64$ in Table 2
(S-mesh) is $5.300e-3$, in Table 3 (V-mesh) is $1.500e-3$ and in
Table 4 (B-mesh) is $8.491e-4$, while at the Step2 the
corresponding errors are $1.298e-5$, $7.363e-7$, $3.313e-7$.
Therefore:

(1) the TGAs significantly increase the accuracy and the
experiments confirm Theorem 2 and Corollary \ref{corollary1};

(2) it is known (see \cite{cite7}) that the most accurate is the
B-mesh and now for the TGAs the situation is similar.

Finally, the CPU time (boldface numbers) is given in Table 4. For
example for $\varepsilon =10^{-1}$, one must compare the value
$0.1406$ with $0.0938$, $1.2344$ with $0.1875$, $34.9688$ with
$8.8906 $. The computational cost of the TGAs is significant and
decreases with $N$.

Table 5 presents results for the two-grid algorithm 2 (TGAs2). One
solves the nonlinear problem (\ref{eq1}) on the coarse grid
($N=4$) and after this (\ref{eq1}) is linearized about the
interpolant of the numerical solution. Then the lirnearized
problem is solved on the fine mesh with $n=N^2=16$. Next, again
the problem (\ref{eq1}) is linearized, but about the interpolant
of the last numerical solution and the obtained linearized problem
is solved on the fine mesh with $n=N^4=256$. Therefore, this
procedure is equivalent to solving the problem (\ref{eq1}) on a
coarse grid with $N=16$ and then the corresponding linearized
problem on a fine mesh with $n=256$. The advantage of the TGAs2 is
in decreasing of the number of the algebraic equations of the
nonlinear difference problem.

\begin{table}[tb]\label{tab2}
\caption{(Example 1) The maximum error and the numerical orders of
convergence for $\varepsilon = 10^{-1},\;10^{-2},\;10^{-4}$ for
the scheme (\ref{eq8}) and the two-grid algorithm 1 (TGAs1) on
S-mesh.}
\begin{tabular}{c c c c c c c}
\hline\noalign{\smallskip}
\hline $\varepsilon $ & N & 8 & 16& 32& 64 \\
\cline{2-6}  & n & 64 & 256 & 1024  &  4096  \\
\noalign{\smallskip}\hline\noalign{\smallskip}
\hline $10^{-1}$&Step 1 & 3,230E-02 & 7,500E-3 & 1,900E-3 & 4.682E-4 \\
\cline{2-6} &$O_{N}$ & 1,4602 & 1,9809 & 2,0209 & \\
\cline{2-6} &Step 2 & 8,295E-4 &  4,450E-5 &  2,844E-6 & 1,763E-7 \\
\cline{2-6} & $O_{n}$& 4,2204 & 3,9679 & 4,0117 & \\
\hline $10^{-2}$&Step 1& 7,460E-2 & 3,920E-2 & 1,480E-2 & 5,300E-3 \\
\cline{2-6} &$O_{N}$ &0,9283 & 1,4053 & 1,4815 & \\
\cline{2-6} &Step 2& 7,000E-3 & 1,100E-3 & 1,299E-4 & 1,298E-5 \\
\cline{2-6} & $O_{n}$& 2,6699 & 3,0823 & 3,3234 & \\
\hline $10^{-4}$&Step 1t& 7,460E-2 & 3,920E-2 & 1,480E-2 & 5,300E-3 \\
\cline{2-6} &$O_{N}$& 0,9283 & 1,4053 & 1,4815  & \\
\cline{2-6} &Step 2& 7,000E-3 & 1,100E-3 & 1,299E-4 & 1,298E-5 \\
\cline{2-6} &$O_{n}$& 2,6699 & 3,0821 & 3,3234 & \\
\noalign{\smallskip}\hline
\end{tabular}
\end{table}

\begin{table}[tb]\label{tab3}
\caption{(Example 1) The maximum error and the numerical orders of
convergence for $\varepsilon = 10^{-1},\;10^{-2}, \;10^{-4}$ for
the scheme (\ref{eq8}) and the (TGAs1) on V-mesh, $a=1,\;q=0.4$.}
\begin{tabular}{c c c c c c c}
\hline\noalign{\smallskip}
\hline $\varepsilon $ & N & 8 & 16& 32& 64 \\
\cline{2-6}  & n & 64 & 256 & 1024  &  4096  \\
\noalign{\smallskip}\hline\noalign{\smallskip}
\hline $10^{-1}$&Step 1 & 2,790E-2 &  8,000E-3 & 2,000E-3 & 5,177E-4 \\
\cline{2-6} &$O_{N}$ & 1,8022 & 2,0000 & 1,9498  & \\
\cline{2-6} &Step 2& 6,929E-4 &  4,647E-5 &  2,942E-6 &  1,873E-7 \\
\cline{2-6} &$O_{n}$ & 3,8985 & 3,9812 & 3,9738   & \\
\hline $10^{-2}$&Step 1& 1,368E-1 &  2,090E-2 &  5,900E-3 & 1,500E-3 \\
\cline{2-6} &$O_{N}$ &2,7105 & 1,8247 & 1,9758 & \\
\cline{2-6} &Step 2& 4,600E-03 &  2,051E-4 &  1,246E-5 &  7,363E-7 \\
\cline{2-6} & $O_{n}$& 4,4873 &  4,0404 & 4,0813 & \\
\hline $10^{-4}$&Step 1& 1,493E-1 &  2,220E-2 &  5,900E-3 & 1,500E-3 \\
\cline{2-6} &$O_{N}$& 2,7496 & 1,9118 & 1,9758  & \\
\cline{2-6} &Step 2& 5,400E-3 &  2,170E-4 &  1,248E-5 &  7,363E-7  \\
\cline{2-6} &$O_{n}$& 4,6369 & 4,1207 & 4,0827 & \\
\noalign{\smallskip}\hline
\end{tabular}
\end{table}

\begin{table}[tb]\label{tab4}
\caption{(Example 1) The maximum error and the numerical orders of
convergence for $\varepsilon = 10^{-1},\;10^{-2},\;10^{-4}$ for
the scheme (\ref{eq8}) and the (TGAs1) on  B-mesh, $a=4,\;q=0.4$.}
\begin{tabular}{c c c c c c c c c}
\hline\noalign{\smallskip}
\hline $\varepsilon $ & N & 8 & 16& 32& 64 & 256 & 1024 \\
\cline{2-8}  & n & 64 & 256 & 1024  &  4096 & - &  - \\
\noalign{\smallskip}\hline\noalign{\smallskip}
\hline $10^{-1}$&Step 1 & 3,230E-2 & 7,500E-3 & 1,900E-3 &
4,682E-4
& 2,925E-5 & 1,8281e-6 \\
\cline{2-8} & $O_{N}$& 2,1066 & 1,9809 & 2,0209 & 2,0005 &  &  \\
\cline{2-8}& CPU & 0.0469 &  0.0625 & 0.0781 & \textbf{0.1406} & \textbf{1.2344} & \textbf{34.9688} \\
\cline{2-8} &Step 2& 8,295E-4 & 4,450E-5 & 2,844E-6 & 1,763E-7 &  &  \\
\cline{2-8} & $O_{n}$& 4,2204 & 3,9679 & 4,0117 & &  &  \\
\cline{2-8}& CPU & \textbf{0.0938} & \textbf{0.1875}  & \textbf{0.6094} & 8.8906 &  &  \\
\hline $10^{-2}$&Step 1& 5,810E-2 & 1,380E-2 & 3,400E-3 & 8,491E-4
&
5,309E-5 & 3,3181e-6 \\
\cline{2-8} & $O_{N}$&2,0739 & 2,0211 & 2,0016 & 1,9994 &  &  \\
\cline{2-8}& CPU & 0.0313 &  0.0469 & 0.0938 & \textbf{0.1406} & \textbf{1.2656} & \textbf{34.7500} \\
\cline{2-8} &Step 2& 1,700E-3 & 8,720E-5 & 5,323E-6 & 3,317E-7 &  &  \\
\cline{2-8} &$O_{n}$ & 4,2850 & 4,0340 & 4,0044 & &  & \\
\cline{2-8}& CPU & \textbf{0.1094} &  \textbf{0.1719} & \textbf{0.5781} & 7.7344 &  &  \\
\hline $10^{-4}$&Step 1& 5,810E-2 & 1,380E-2 & 3,400E-3 & 8,491E-4 & 5,309E-5 & 3,3181e-6 \\
\cline{2-8} &$O_{N}$& 2,0739 & 2,0211 & 2,0016 & 1,9994 &  &  \\
\cline{2-8}& CPU & 0.0581 & 0.0781  & 0.0938 & \textbf{0.1494} & \textbf{1.2500} & \textbf{34.8594} \\
\cline{2-8} &Step 2& 1,700E-3 & 8,720E-5 & 5,323E-6 & 3,317E-7 &  &  \\
\cline{2-8} &$O_{n}$& 4,2850 & 4,0340 & 4,0044 & &  &  \\
\cline{2-8}& CPU & \textbf{0.1094} & \textbf{0.2031}  & \textbf{0.5625} & 6.9844 &  &  \\
\noalign{\smallskip}\hline
\end{tabular}
\end{table}

\begin{table}[tb]\label{tab8}
\caption{(Example 1) The maximum error and the numerical orders of
convergence for $\varepsilon = 10^{-1},\;10^{-2}, \;10^{-4}$ for
the scheme (\ref{eq8}) and the (TGAs2) on V-mesh, $a=3,\;q=0.4$.}
\begin{tabular}{c c c c c c}
\hline \noalign{\smallskip}
\hline $\varepsilon $ & N & 4 & 8 & 16 \\
\cline{2-5}  & n & 16 & 64 & 256 \\
\cline{2-5}  & $n^2$ & 256 & 4096 & 65536 \\
\noalign{\smallskip}\hline\noalign{\smallskip}
\hline $10^{-1}$&Step 1 & 1.094E-1 &  2.780E-2 & 7.100E-3 \\
\cline{2-5} &$O_{N}$ & 1.9765 & 1.9692 & \\
\cline{2-5} &Step 2& 2.412E-3 &  1.555E-4 &  9.744E-6 \\
\cline{2-5} &$O_{n}$ & 3.9552 & 3.9963 & \\
\cline{2-5} &Step 3& 6.729E-6 &  2.759E-8 &  1.113E-10 \\
\cline{2-5} &$O_{n^2}$ & 7.9299 & 7.9533 & \\
\hline $10^{-2}$&Step 1& 2.163E-1 &  5.480E-2 &  1.390E-2 \\
\cline{2-5} &$O_{N}$ &1.9808  & 1.9791 &  \\
\cline{2-5} &Step 2& 5.280E-2 &  3.500E-3 &  2.446E-4 \\
\cline{2-5} & $O_{n}$& 3.9151 &  3.8389 &  \\
\cline{2-5} &Step 3& 1.088E-3 & 5.094E-6& 2.292E-8 \\
\cline{2-5} &$O_{n^2}$ &7.3213  & 7.7918 & \\
\hline $10^{-4}$&Step 1& 4.334E-1 &  1.103E-1 &  2.780E-2 \\
\cline{2-5} &$O_{N}$& 1.9743 & 1.9883 & \\
\cline{2-5} &Step 2& 3.887E-2  &  3.502E-3 &  2.546E-4   \\
\cline{2-5} &$O_{n}$& 3.4724 & 3.7819 &  \\
\cline{2-5} &Step 3& 8.016E-4  &  5.225E-6 & 2.301E-8  \\
\cline{2-5} &$O_{n^2}$ & 7.2614 & 7.8267 & \\
\noalign{\smallskip}\hline
\end{tabular}
\end{table}

\noindent \textbf{Example 2.} \label{exa2}
The theoretical results in Sections 3, 4 concern the classical
case of singularly perturbed reaction problems. We will
demonstrate by this example the efficiency of the TGAs applied to
another class of problems. We consider the test problem in
\cite{cite19}:
$$ -\varepsilon ^2 \left(\frac{u'}{u+1}\right)' +u=f(x), \;
u(0)=1,u(1)=\beta.$$

The method discussed in \cite{cite19} is the central
finite-difference scheme applied on meshes of Bakhvalov and
piecewise-equidistant types. Here we used a similar
discretization.

We chose $f(x)$ and $\beta$ so that the exact solution is
$$ u_{\varepsilon}(x)=\exp(-x/\varepsilon)+\exp(x)-1.$$

\noindent Now the solution has a boundary layer at $x=0 $. The
results in Tables 6, 7, 8 are similar to those of Example 1. In Table 6 we also give experimental results
concerning an optimal choice of the number $r$ based on the
approximate (up to constant multiplier) relation
$\frac{N^r}{r}\approx \frac{N^2}{\ln N}$. The accuracy is lower
and the convergence is slower but there is a decrease in the
computational cost.

\begin{table}\label{tab6}
\caption{(Example 2) The maximum error and the numerical orders of
convergence for $\varepsilon = 10^{-1},\;10^{-2},\;10^{-4}$ for
the scheme (\ref{eq8}) and the (TGAs1) on S-mesh.}
\begin{tabular}{c c c c c c c}
\hline\noalign{\smallskip}
\hline $\varepsilon $ & N & 8 & 16& 32& 64 \\
\cline{2-6}  & n & 64 & 256 & 1024  &  4096  \\
\cline{2-6}  & r & 1.9752 & 1.8550 & 1.8131  &  1.7984  \\
\cline{2-6}  & $N^r$ & \textbf{60} & \textbf{172} & \textbf{536}  &  \textbf{1772}  \\
\noalign{\smallskip}\hline\noalign{\smallskip}
\hline $10^{-1}$&Step 1 & 4,424E-03 &  1,079E-03 & 2,677E-04 & 6,678E-05 \\
\cline{2-6} &$O_{N}$ & 2,0109 & 2,0109 & 2,0029 & \\
\cline{2-6} &Step 2& 6,248E-04&  8,443E-05 &  1,454E-05  &  1,472E-06 \\
\cline{2-6} &$O_{n}$ & 2,8877 & 2,5376 & 3,3048   & \\
\cline{2-6} &Step 2& 1,550E-03&  2,086E-04 &  2,575E-05  &  3,010E-06 \\
\cline{2-6} &\textbf{$O_{N^r}$} & \textbf{2,8931} & \textbf{3,0180} & \textbf{3,0968}   & \\
\hline $10^{-2}$&Step 1& 8,790E-03 &  3,834E-03 & 1,477E-03 & 5,255E-04 \\
\cline{2-6} &$O_{N}$ &1,1970 & 1,3762 &1,4909 & \\
\cline{2-6} &Step 2& 3,001E-03 &  8,396E-04 &   1,582E-04 &  2,094E-05 \\
\cline{2-6} & $O_{n}$& 1,8374 &  2,4079 & 2,9175 & \\
\cline{2-6} &Step 2& 5,322E-03&  1,891E-03 &  4,704E-04  &  9,902E-05 \\
\cline{2-6} &$O_{N^r}$ & \textbf{1,4309} & \textbf{2,0071} & \textbf{2,2482}   & \\
\hline $10^{-4}$&Step 1& 3.128E-2 &  1.085E-2 &  2.683E-3 & 6.695E-4 \\
\cline{2-6} &$O_{N}$& 1.5279 & 2.0153 & 2.0031  & \\
\cline{2-6} &Step 2& 1.935E-3 &  1.168E-4 &  7.087E-6 &  4.444E-7  \\
\cline{2-6} &$O_{n}$& 4.0505 & 4.0427 & 3.9953 & \\
\cline{2-6} &Step 2& 5,293E-03&  1,888E-03 &  4,708E-04  &  9,912E-05 \\
\cline{2-6} &$O_{N^r}$ & \textbf{1,4871} & \textbf{2,0037} & \textbf{2,2479}   & \\
\noalign{\smallskip}\hline
\end{tabular}
\end{table}

\begin{table}\label{tab5}
\caption{(Example 2) The maximum error and the numerical orders of
convergence for $\varepsilon = 10^{-1},\;10^{-2},\;10^{-4}$ for
the scheme (\ref{eq8}) and the (TGAs1) on V-mesh, $a=2,\;q=0.4$.}
\begin{tabular}{c c c c c c c}
\hline\noalign{\smallskip}
\hline $\varepsilon $ & N & 8 & 16& 32& 64 \\
\cline{2-6}  & n & 64 & 256 & 1024  &  4096  \\
\noalign{\smallskip}\hline\noalign{\smallskip}
\hline $10^{-1}$&Step 1 & 8,117E-3 &  1,835E-3 & 5.333e-4 & 1.331e-4 \\
\cline{2-6} &$O_{N}$ & 2.1699 & 1.7550 & 2.0020  & \\
\cline{2-6} &Step 2& 4.470e-4&  8.554e-5 &  6.366e-6 &  4.387e-7 \\
\cline{2-6} &$O_{n}$ & 2.3855 & 3.7481 & 3.8589   & \\
\hline $10^{-2}$&Step 1& 3.079E-2 &  1.098E-2 &  2.718E-3 & 6.775E-4 \\
\cline{2-6} &$O_{N}$ &1.4873 & 2.0150 & 2.0040 & \\
\cline{2-6} &Step 2& 2.056E-3 &  1.148E-4 &  6.971E-6 &  4.368E-7 \\
\cline{2-6} & $O_{n}$& 4.1623 &  4.0422 & 3.9964 & \\
\hline $10^{-4}$&Step 1& 3.128E-2 &  1.085E-2 &  2.683E-3 & 6.695E-4 \\
\cline{2-6} &$O_{N}$& 1.5279 & 2.0153 & 2.0031  & \\
\cline{2-6} &Step 2& 1.935E-3 &  1.168E-4 &  7.087E-6 &  4.444E-7  \\
\cline{2-6} &$O_{n}$& 4.0505 & 4.0427 & 3.9953 & \\
\noalign{\smallskip}\hline
\end{tabular}
\end{table}

\begin{table}[ht]\label{tab7}
\caption{(Example 2) The maximum error and the numerical orders of
convergence for $\varepsilon = 10^{-1},\;10^{-2},\;10^{-4}$ for
the scheme (\ref{eq8}) and the (TGAs1) on B-mesh, $a=2,\;q=0.4$.}
\begin{tabular}{c c c c c c c}
\hline\noalign{\smallskip}
\hline $\varepsilon $ & N & 8 & 16& 32& 64 \\
\cline{2-6}  & n & 64 & 256 & 1024  &  4096  \\
\noalign{\smallskip}\hline\noalign{\smallskip}
\hline $10^{-1}$&Step 1 & 8,668E-3 &  2,197E-3 & 5,593E-4 & 1,413E-4 \\
\cline{2-6} &$O_{N}$ & 1,9802 & 1,9737 & 1,9850 & \\
\cline{2-6} &Step 2& 2,627E-4&  1,776E-5 &  1,115E-6 &  7,353E-8 \\
\cline{2-6} &$O_{n}$ & 3,8863 & 3,9944 & 3,9220   & \\
\hline $10^{-2}$&Step 1& 2,605E-2 &  5,828E-3 &  1,218E-3 & 2,987E-4 \\
\cline{2-6} &$O_{N}$ &2.1602 & 2.2587 & 2.0274 & \\
\cline{2-6} &Step 2& 5.069E-4 &  3.051E-5 &  1.677E-6 &  1.032E-7 \\
\cline{2-6} & $O_{n}$& 4.0544 &  4.1856 & 4.0216 & \\
\hline $10^{-4}$&Step 1& 4.038E-2 &  1.077E-2 &  2.666E-3 & 6.625E-4 \\
\cline{2-6} &$O_{N}$& 1.9065 & 2.0146 & 2.0528  & \\
\cline{2-6} &Step 2& 8.857E-4 &  6.401E-5 &  4.009E-6 &  2.401E-7  \\
\cline{2-6} &$O_{n}$& 3.7905 & 3.9970 & 4.0611 & \\
\noalign{\smallskip}\hline
\end{tabular}
\end{table}

\section{Conclusions and future work } In this paper, we proposed two-grid
algorithms for the finite difference solution of singularly
perturbed reaction-diffusion problems. In these two-grid
algorithms, the solution of the fully nonlinear \textit{coarse}
problem is used in a single-step \textit{linear} fine mesh
problem. Numerical experiments demonstrate that the two-grid
algorithms are \textit{dramatically more efficient} than the
standard one-grid algorithm.

We have studied the semilinear Dirichlet boundary value problem
(\ref{eq1})-(\ref{eq3}). It is clear that the present approach can
be easily extended to  equation (\ref{eq1}) with any set of linear
two-point boundary conditions \cite{cite1,cite2}. Our future work
will be devoted to the development of the proposed algorithms for
convection-dominated equations and systems of equations in one and
two dimensions, for example the models in finance \cite{K13, K14, K16} and medicine \cite{K15}. Results concerning the two-grid algorithms
combined with exponential difference schemes on standard meshes
for singularly perturbed nonlinear ordinary differential equations
and systems of equations were reported in \cite{cite20,cite21}.\\

\textbf{Acknowledgements.} The authors would like to thank the
referees for their criticism and valuable comments. Also, thanks
to Prof. N. Kopteva for careful reading the last version of the
manuscript and for the suggestions to improve the readability of
the paper.\\

This research was supported by the Bulgarian National Fund of
Science under Projects Bg-Sk-203 and ID$\_ $ 09$\_ $ 0186.

\bibliographystyle{spmpsci}

\begin{thebibliography}{25}

\bibitem{cite24} V. B. Andreev, On the uniform convergence of a
classical difference scheme on a nonuniform mesh for the
one-dimensional dingularly perturbed reaction-diffusion equation,
Zh. Vychisl. Mat. i Mat. Fiz., 44 (2004), 476492 (in Russian),
translation in Comput. Math. Math. Phys., 44, 449--464 (2004)

\bibitem{cite1} I. Tr. Angelova, L.G. Vulkov: Comparison of the two-grid method on different meshes for a
singularly perturbed semilinear problem,Amer. Inst. of Phys. 1067,
305-312 (2008)

\bibitem{cite2} U. M. Asher, R.M.Matheij, R.D.Russel, Numerical Solution of Boundary Value Problems for
Ordinary Differential Equations, SIAM, Philadelphia, 1995

\bibitem{cite3} O. Axelsson, On mesh independence and Newton methods, \emph{Applications of Math.}, No 4-5, 249--265 (1993)

\bibitem{cite4} N.S. Bakhvalov: On the optimization methods for solcing boundary value problems with
bounsary layers, Zh. Vychisl. Math. Fiz. 24, 841--859 (1969)(in
Russian)

\bibitem{cite5} R.E. Bellman, R.E. Kalaba, Quasilinearization and Nonlinear Boundary-Value Problems,
Elsevier Publishing Company, New York (1965)

\bibitem{cite6} P.A. Farrell, J.J.H. Miller, E. O'Riordan, and G.I.
Shishkin, On the non-existence of $\varepsilon$-uniform finite
difference method on uniform meshes for semilinear two-point
boundary value problems, \emph{Math. Comp., 67}, N 222, 603--617
(1999)

\bibitem{cite7} D.~Herceg, K.~Surla, I.Radeka, and H.~Mali\v{c}i\'c, Numerical experiments with different schemes for singularly
 perturbed problem, \emph{Novi Sad J. Math. 31}, 93-101 (2001)


\bibitem{K13}M.N. Koleva, Efficient numerical method for solving Cauchy problem for the Gamma equation, AIP CP 1410 (2011) 120--127.
\bibitem{K14}M.N. Koleva, Iterative methods for solving nonlinear parabolic problem in pension saving management, AIP CP 1404 (2011) 457--463.

\bibitem{K15}	M.N. Koleva, Efficient numerical method for solving 1D degenerate Keller-Segel systems, AIP CP 1497 (2012) 168--175. 

 \bibitem{K16} M.N. Koleva, Positivity preserving numerical method for non-linear Black-Scholes models, LNCS 8236 (2013) 363--370.
 
 
\bibitem{cite25} N.~Kopteva, Maximum norm a posteriori error estimates
for a one-dimensional singularly perturbed semilinear
reaction-diffusion problem, IMA J. Numer. Anal., 27, 576--592
(2007)

\bibitem{cite8} N.~Kopteva, and M.~Stynes,
Numerical analysis of singularly perturbed nonlinear
reaction-diffusion problems with multiple solutions, \emph{Comput.
Math. Appl. 51}, 857--864 (2006)

\bibitem{cite9} V. Lakshmikantham, A.S. Vatsala, Generalized Quasilinearization for Nonlinear Problems, Mathematics and its Applications,
 Vol. 440, Kluwer Academic Publishers, Dordrecht (1998)


\bibitem{cite10} T. Linss, Layer-Adapted Meshes for Convection-Diffusion Problems,
Lect. Notes in Math, v. 1985, Springer, Heidelberg, Berlin (2010)

\bibitem{cite11} J.J.H. Miller, E. O'Riordan, and G.I.
Shishkin, Fitted Numerical Methods for Sinhular Perturbed
Problems, \emph{World Scientific, Singapore}(1996)

\bibitem{cite12} M.~Mu, and J.~Xu,
A two-grid method of mixed Stokes-Darcy model for coupling fluid
flow with porous media flow, \emph{SIAM J. Numer. Anal.} {45},
1801--1813 (2007)

\bibitem{cite13} H. Roos, M. Stynes, and L. Tobiska, Numerical
Methods for Singular Perturbed Differential Equations,
Convection-Diffusion and Flow Problems, \emph{Springer-Berlin}
(2008)

\bibitem{cite25} I. A. Savin, On the uniform convergence with
respect to a  small parameter of difference schemes for an
ordindry differential equation. \emph{Zh. Vychisl. Mat. i Mat.
Fiz.}, 35, 1758-1765(1995)(in Russian)

\bibitem{cite14} G.I. Shishkin, L.P. Shishkina, A high-order
Richardson method for a quasilinear singularly perturbed elliptic
reaction-diffusion equation, \emph{Diff. Eqns}, v.41, No 7,
1030--1039 (2005)

\bibitem{cite15} G. Sun, and M. Stynes,
An almost fourth order uniformly convergent difference scheme for
semilinear singularly perturbed reaction-diffusion problem,
\emph{Numer. Math., 7a}, 487--500 (1995)

\bibitem{cite16} R.~Vulanovi\'c,
Mesh Construction for Discretization of Singularly Perturbed
Boundary Value Problems, Ph. D. Thesis, Univ. of Novi Sad, (1986)

\bibitem{cite17} R. Vulanovi\'c. Fourth order algorithms for a semilinear singular perturbation
problem, Numer. Algorithms 16, 117--128 (1997)

\bibitem{cite18} R. Vulanovi\'c, An almost sixth-order
finite-difference method for semilinear singular perturbation
problems, Comput. Methods Appl. Math., 4, 368--383 (2004)


\bibitem{cite19} R. Vulanovi\'c, Finite-difference methods for a class to strongly nonlinear
singular perturbed problems, Num. Math. Theor. Appl., v.1, No 2,
235--244 (2008)

\bibitem{cite20} L. Vulkov, A. Zadorin, Two-grid interpolation
algorithms for difference schemes of exponential type for
semilinear diffusion convection-dominated equations \emph{AIP CP,
1067}, 284--292 (2008)

\bibitem{cite21} L. G. Vulkov, A. Zadorin: A Two-grid algorithm for solution of the difference equations of a
system of singularly perturbed semilinear equations, LNCS
Springer, v.5434, 582--589 (2009)

\bibitem{cite22} A. Zadorin, Method of interpolation for a boundary layer problem, Sib. J. of Numer. Math.,
v.10, N3, 267--275 (2007)(in Russian)

\bibitem{cite23} J. Xu, A novel two-grid method for semilinear ellipptic equations, \emph{SIAM}, J. Sci. Comput.,v. 15, No 1, 231--237 (1994)

\end{thebibliography}

\end{document}